\documentclass[aps,prb,twocolumn,superscriptaddress,floatfix]{revtex4}
\usepackage{amsfonts}
\usepackage{amsmath}
\usepackage{amssymb}
\usepackage{graphicx}

\begin{document}

\title{Current induced magnetization reversal in SrRuO$_{3}$}
\author{Yishai Shperber}
\affiliation{Department of Physics, Nano-magnetism Research
Center, Institute of Nanotechnology and Advanced Materials,
Bar-Ilan
University, Ramat-Gan 52900, Israel}
\author{Daniel Bedau}
\affiliation{Department of Physics, New York University, New York, New York 10003 USA}
\author{James W. Reiner}
\affiliation{HGST, 3403 Yerba Buena Rd, San Jose, CA 95315 USA}
\author{Lior Klein}
\affiliation{Department of Physics, Nano-magnetism Research
Center, Institute of Nanotechnology and Advanced Materials,
Bar-Ilan
University, Ramat-Gan 52900, Israel}

\begin{abstract}
We inject current pulses into uniformly magnetized patterns of thin films of the itinerant ferromagnet SrRuO$_{3}$, while monitoring the
  effective temperature of the patterns during the current injection. We gradually increase the amplitude of the pulses  until magnetization reversal
  occurs.
  We observe magnetization reversal induced by current above a temperature dependent threshold and show that this effect is not simply due to sample heating or Oersted fields.
 We discuss the applicability of current-induced spin-wave instability scenario.
\end{abstract}






\maketitle

\section{introduction}
The shrinking size of spintronics devices and the need for efficient and scalable methods for their manipulation has led to enhanced interest in spin-torque effects of electrical current on the magnetic configuration of nanostructures.
The study in this direction has focused so far on effects related to magnetic nonuniformity; in particular, the effect of current on ferromagnetic domain walls which yields domain wall motion \cite{Berger1984,TataraDWM2004,Li2004,Yamaguchi2004,Feigenson2007,Boulle2008} and the effect of current on magnetic heterostructures which yields magnetic switching \cite{Berger1996,Slonczewski1996,Tsoi1998,Myers1999,Katine2000}. Both effects have been observed in different systems and they appear to be useful for novel memory devices. A more subtle effect of magnetic nonuniformity is expected when a single uniformly magnetized nanostructure is connected via asymmetric contacts to normal metals. In this case, asymmetric spin accumulation near the two contacts may induce magnetic instability \cite{Polianski2004,Stiles2004,Ji2003,Ozilmaz2004}.

Based on several theoretical works, electrical current is expected to induce magnetization reversal also in a uniformly magnetized ferromagnet irrespective of the contact geometry \cite{Slonczewski1999,TataraNuc2005,Li2005,Korenblit2008} due to a current-induced spin wave instability. This is a fundamental phenomenon with practical implications on the  functionality of spintronic devices; hence, its experimental study is of particular importance for the field of spintronics. Nevertheless,  so far the study of this  phenomenon has been very sparse \cite{Togawa2008,Feigenson2008,Gerber2010}.

Current-induced magnetization reversal in uniformly magnetized films is expected at relatively high current densities. Such currents generate heat and high Oersted-fields which may also induce magnetization reversal. Therefore, reliable identification of the phenomenon requires the ability to isolate the effect of the current itself.

Here we use the itinerant ferromagnet SrRuO$_{3}$  \cite{RevSrRuO3}, a material characterized by large spin polarization of its conducting electrons \cite{Ray2003}. This material exhibits large domain wall resistivity \cite{Klein2000} and efficient current-induced domain wall motion \cite{Feigenson2007} indicating strong effect of the current on the magnetic configuration and vice versa.

In this work, we observe magnetization reversal induced by current above a temperature dependent threshold and show that this effect is not simply due to sample heating or Oersted fields. Our observations are obtained for a wide range of temperatures while applying various fields including a zero field and fields suppressing reversal. Basic features of the results are consistent with the spin wave instability scenario \cite{Li2005} suggesting that this is the relevant mechanism.

\begin{figure}[b]
\includegraphics[scale=0.45, trim=50 230 50 270]{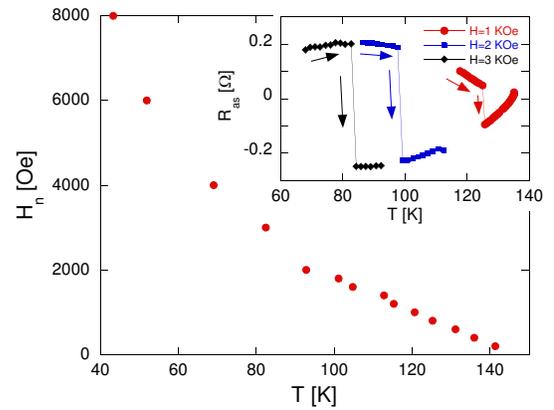}
\caption{The magnetization reversal field vs. temperature in the absence of current effects (The measureing current is 30 $\mu$A). Inset: The antisymmetric transverse resistance vs. temperature with reversing fields between 1 KOe and 3 KOe. The sharp jumps are attributed to magnetization reversal.
} \label{1}
\end{figure}

\section{experimental details}
Our samples are high-quality epitaxial films of SrRuO$_{3}$ which were
 grown on slightly
miscut (2$^{\circ}$) SrTiO$_{3}$ substrates.
The films are orthorhombic
($a=5.53$ \AA , $b=5.57$ \AA , $c=7.85$ \AA) with the
 $c$ axis in the film plane (perpendicular to the miscut direction) and the $a$ and $b$
axes are at 45$^{\circ}$ relative to the film plane.
The Curie temperature of the films is $\sim150$ K and they exhibit
uniaxial magnetocrystalline anisotropy with the easy axis changing in the (001) plane between
the $b$ axis at $T\geq$$T_{c}$ to $30^\circ$ from the film normal at low temperatures \cite{Marshall}. The ratio between the resistivity at 300 K and the resistivity at the low temperature limit is greater than ten.  The films are patterned for magnetotransport measurements using e-beam lithography and $Ar^+$ ion milling with a typical current path width of 1.5 $\mu$m. The data presented below are for film thickness of 37.5 nm.

The magnetization of the patterns is monitored by measuring
the Hall effect (HE) which consists (as in other magnetic conductors) of an ordinary Hall effect (OHE) determined by the perpendicular component of the magnetic field and an anomalous Hall effect (AHE) related to the perpendicular component of the magnetization. Since the easy axis for magnetization in SrRuO$_3$ is tilted out of the plane there is a contribution of the AHE in the absence of an external field \cite{Klein2000,Fang2003}.

\section{results and discussion}
As we study the contribution of current to magnetization reversal with and without an applied magnetic field, we first determine the temperature dependence of the reversal field, $H_n$, in the absence of current effects \cite{Genish2004}. In these measurements we monitor the magnetization with a small current of 30 $\mu$A  (current density $J\sim 4.7\times 10^{4}\frac{\rm{A}}{\rm{cm}^{2}}$) which is too small to contribute to the magnetization reversal (see Figure~\ref{1}). We fully magnetize the sample at a low temperature and then apply a certain reversing field and warm the sample at a rate of 20 K/min until reversal occurs (see inset of Figure~\ref{1}). In this experiment and in the experiments described below the magnetic field is applied perpendicular to the film plane.
%

To explore current contribution to magnetization reversal, we fully magnetize the sample at an initial temperature, $T_b$, between 28 and 145 K and apply a current pulse of constant duration. The amplitude of the pulse is gradually increased and the magnetization is monitored after each pulse by measuring the AHE with a probing current of 30 $\mu$A.
We repeat these experiments with various magnetic fields, both assisting and suppressing magnetization reversal.

As we show below, our main observation is that when the amplitude of the pulse exceeds a certain value the pattern is no longer fully magnetized.

\begin{figure}[t]
\includegraphics[scale=0.45, trim=50 0 50 50]{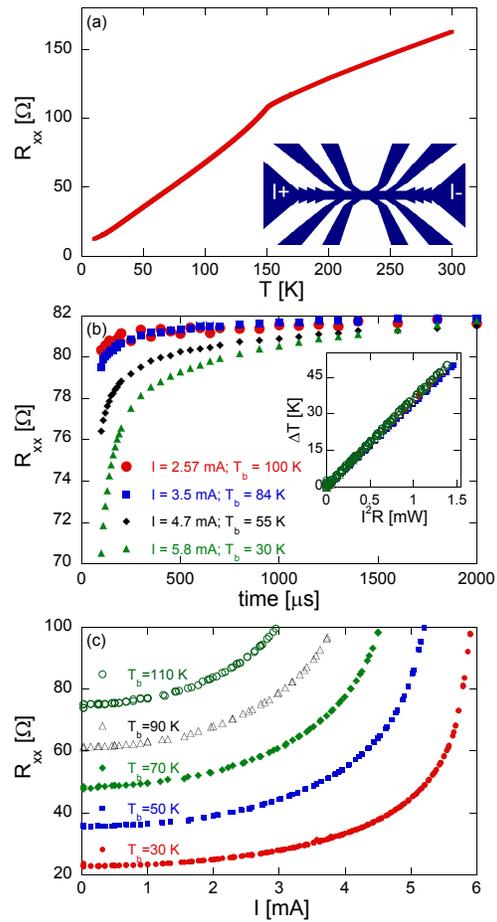}
\caption{(a) The longitudinal resistance measured between two neighboring voltage leads vs. temperature. Inset: A scheme of the pattern. (b) The longitudinal resistance ($R_{xx}$) vs. pulse duration. The pulse amplitude varies between 2.57 and 5.8 mA, and the initial temperature varies between 100 and 30 K, respectively. $R_{xx}$ is measured during the last 50 $\mu$s of the pulse. Inset: The change in the local temperature vs. the current's power. The pulse duration is 1 ms. (c) The longitudinal resistance during the last 50 $\mu$s of the pulse vs. the pulse amplitude for different initial temperatures. The total pulse duration is 1 ms.
} \label{2}
\end{figure}

The currents that induce magnetic instability correspond to high current densities (1 mA corresponds to current density of $\sim 1.77\times10^{6}\frac{\rm{A}}{\rm{cm}^{2}}$); therefore, the effect of heating should be carefully considered.
To identify changes in the temperature of the pattern during the pulse, we use the measured temperature dependence of the resistance of the pattern as our calibrated temperature sensor; see the temperature dependence of the longitudinal resistance, $R_{xx}$, measured
between two neighboring voltage leads (Figure~\ref{2}(a)).

Figure~\ref{2}(b) shows changes in $R_{xx}$ during current pulses with various amplitudes. The duration of the pulse is 2 ms and $R_{xx}$ is measured in 50 $\mu$s intervals. The temperature of the pattern before the current pulse is applied, $T_b$, is between 30 K and 100 K and the current amplitude is between 2.57 mA (for $T_b$=100 K) to 5.8 mA (for $T_b$=30 K). We note a fast initial increase in the resistance followed by saturation.

In the current-induced magnetization reversal measurement shown below we use 1 ms pulses with amplitudes which do not exceed 5.84 mA, and we measure $R_{xx}$ during the last 50 $\mu$s of the pulse. In these conditions, the measurement reflects well the highest value of $R_{xx}$ during the pulse.

Figure~\ref{2}(c) shows $R_{xx}$ measured during the last 50 $\mu$s of the pulse as a function of the pulse amplitude. Based on the temperature dependence of $R_{xx}$ (Figure~\ref{2}(a)) we can determine the effective temperature, $T_{\rm{eff}}$, at the end of the current pulse. As this is practically a steady state, we expect that the power dissipation in the pattern, given by $I^2R$, would be balanced by the heat flow.
The inset of Figure~\ref{2}(b) shows a linear dependence between $\Delta T= T_{\rm{eff}} - T_b$ and  $I^2R$ suggesting heat flow proportional to $\Delta T$ with practically temperature-independent thermal conductivity in the relevant temperature regime.

\begin{figure}[t]
\includegraphics[scale=0.4, trim=50 50 50 100]{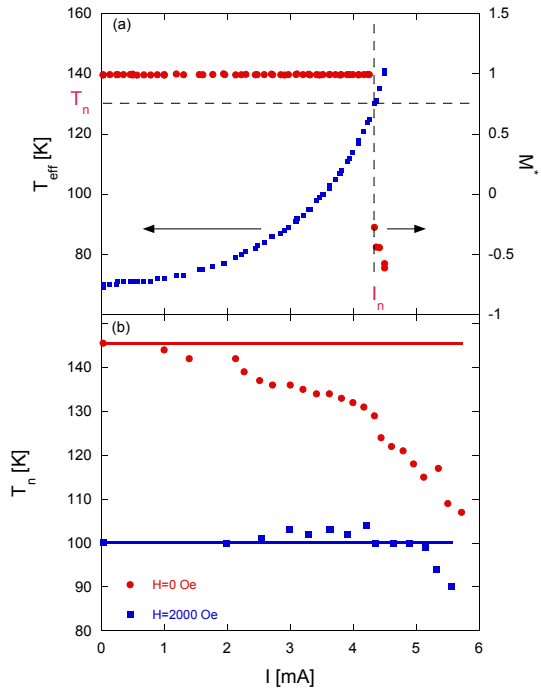}
\caption{(a) The effective temperature (left) and the normalized average magnetization, $M^*$ (right) vs. current amplitude for $T_b$=70 K and a reversing field of $H$=100 Oe. The dashed lines are guide lines for the magnetization reversal current, $I_{n}$, and the magnetization reversal temperature, $T_{n}$. (b) The magnetization reversal temperature, $T_n$, at $H$=0 and 2000 Oe vs. the magnetization reversal current. The solid lines represent the magnetization reversal temperature in the limit of zero current.
} \label{3}
\end{figure}

Figure~\ref{3}(a) shows a typical measurement of current induced magnetization reversal performed at $T_b$=70 K with a reversing field of $H$=100 Oe. The amplitude of the current pulse is gradually increased and the figure shows the effective temperature  $T_{\rm{eff}}$
during the pulse and the normalized average
magnetization, $M^*$, as determined after the pulse at the pre-pulse temperature by measuring the
AHE with a small probing current and dividing the obtained signal by the signal corresponding
to full magnetization (the sample cools down to its pre-pulse temperature in less than 1 second).
Based on such measurements we determine the current, $I_n$,
temperature, $T_n$, and magnetic field, $H_n$, at which
magnetization reversal occurs. The results are insensitive to current polarity.

Figure~\ref{3}(b) presents $T_n$ (the magnetization reversal temperature) as a function of the current pulse amplitude when no magnetic field is applied and when a reversing field of 2000 Oe is applied perpendicular to the film plane. The horizontal lines mark $T_n$ for the two fields in the limit of zero current as determined in temperature sweep experiments (see Figure~\ref{1}). To produce the curves we cool the sample to different temperatures and gradually increase the current pulse. The points represent the current amplitude for which first magnetization reversal occurred and the highest temperature of the sample during the pulse.
For $H$=2000 Oe, $T_n$ is practically independent of $I_n$ for a wide range of current amplitudes, suggesting no current effect beyond the Joule heating which we monitor. On the other hand, for $H$=0, $T_n$ significantly decreases with increasing current, indicating non-trivial current effects.

In addition to heating, current may affect the magnetization reversal due to the Oersted field it induces \cite{Boulle2009}. Based on calculations, we find that the highest field (at the edges of the current path) induced by the largest current we use, 5.8 mA, is $\sim$ 40 Oe normal to the film plane, small compared to $H_n$ at the relevant temperatures.

\begin{figure}[t]
\includegraphics[scale=0.5, trim=00 50 000 50]{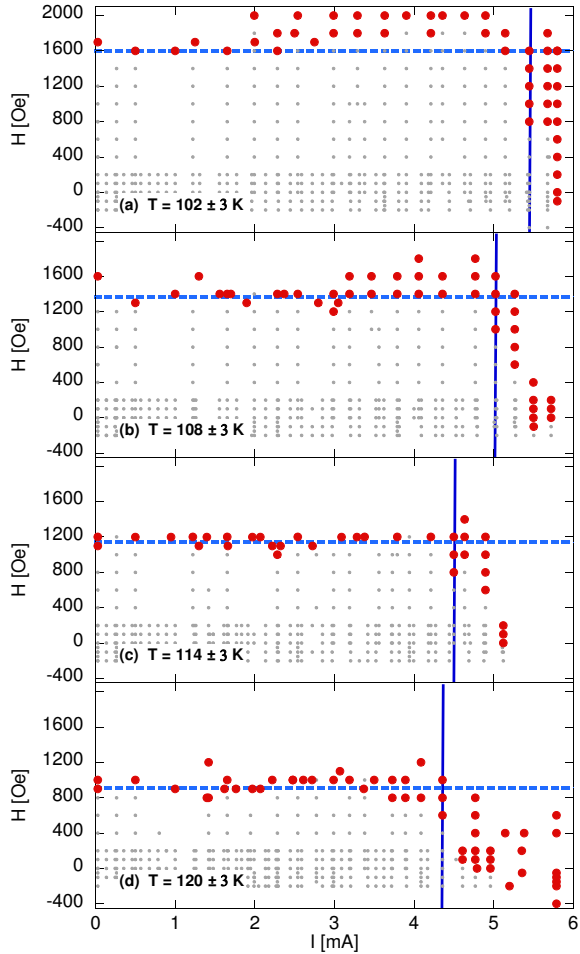}
\caption{Reversal events (big red dots) and non-reversal events (small gray dots) in (I, H) phase space for $T_{\rm{eff}}$ ($\pm$3 K) between 102 and 120 K. The dashed lines represent $H_n$ in the limit of zero current at the upper bound of the temperature regime, and the solid lines represent the threshold currents, $I_t$.
} \label{4}
\end{figure}

Based on measurements as shown in Figure~\ref{3}(a), we construct  Figure~\ref{4} which shows for the vicinity of a given $T_{\rm{eff}}$  (within $\pm3$ K) the combinations of currents and fields that produce first reversal or no-reversal.
Such plots allow us to explore the nonthermal effect of the current as its heating effect is
already taken into consideration. The dashed line represents $H_n$ in the limit of zero current for the upper bound of $T_{\rm{eff}}$.
The solid line marks the threshold current, $I_t$, above which the current affects the magnetization reversal in a non-trivial way. We define this current as the current for which $H_n$ is smaller than its zero limit value by  1.5 times the standard deviation of its distribution for lower currents. The fact that the plot shows multiple values of $H_n$ for a given $I$ has several reasons: the spread in $T_{\rm{eff}}$, which is related to the finite increments of the current pulse, the stochastic nature of the effect, and weak field dependence of $I_n$.
At $T_{\rm{eff}}=102\pm3$ K (Figure~\ref{4}(a)) we identify a sharp transition from a current independent regime to a field independent regime. While for currents lower than $I_t$ the field is current independent, for currents higher than $I_t$
the field dependence is very weak and almost the same current induces magnetization reversal for a wide range of fields including fields which oppose reversal.  As $T_{\rm{eff}}$ increases, the transition becomes more gradual.


\begin{figure}[t]
\includegraphics[scale=0.33, trim=50 150 50 200]{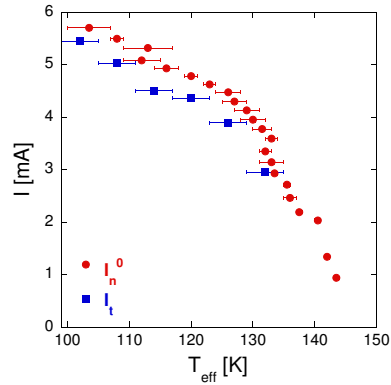}
\caption{The switching current at zero field, $I_n^0$, (red dots) and the threshold current, $I_t$, (blue squares) as a function of temperature.
} \label{5}
\end{figure}

Figure~\ref{5} shows the temperature dependence of $I_t$ and of $I_n^0$ defined as the current at which $H_n$ becomes zero; namely, the current which produces magnetization reversal without external field assistance.  $I_t$ is extracted from Figure~\ref{4}; therefore, $T_{\rm{eff}}$  has an error bar of $\pm3$ K. On the other hand, for  $I_n^0$ the error bar is related to the temperature increment produced by the current increment in our experiments.
We  note that Ref.~\cite{Korenblit2008} suggests a threshold current inversely proportional to the AHE coefficient. In SrRuO$_3$ the AHE vanishes at $T\sim 127$ K; however, we see no divergence of the reversing currents in the vicinity of that temperature.

The existence of a threshold current appears to be consistent with the spin wave scenarios \cite{Slonczewski1999,TataraNuc2005,Li2005}.
Following Ref.~\cite{Li2005}, the threshold current density, $j_{c}$, in the zero-temperature limit is given by $j_{c}=\frac{\gamma eM_{S}l_{w}}{2\pi P\mu_{B}}\sqrt{\frac{H_{K}}{2}}\left (\sqrt{H_{ext}+H_{K}}+\sqrt{H_{ext}+H_{K}+4\pi M_{S}}\right)$, where $\gamma, e$, and $\mu_{B}$ are the gyromagnetic ratio, electron charge and Bohr magneton respectively, $l_{w}$ and $P$ are the domain wall width and the polarization of the current, $M_{S}, H_{K}$ and $H_{ext}$ are the saturation magnetization, the anisotropy field and the external field, respectively.
Since the relevant external fields which we apply are negligible relative to the large anisotropy field of SrRuO$_{3}$ ($>7$ T) \cite{Langner2009,Kats2005}, we expect a weak field dependence of $j_c$, as observed.

Substituting the parameters of SrRuO$_3$ yields at zero temperature current density $\sim8\times10^{8} \frac{\rm{A}}{\rm{cm^2}}$
and a critical wave length of the spin wave excitation of about $\sim$4.2 nm, while our measurements at $T_{\rm{eff}}$ = 102 K yield $\sim10^{7} \frac{\rm{A}}{\rm{cm^2}}$. At least some of the discrepancy may
be attributed to the fact that the model is developed
at zero temperature while our measurements are at
$T>0.7 T_c$. We note that for Co the expected $j_c$ is
more than one order of magnitude higher \cite{Li2006} with a critical wave length of about $\sim$30 nm \cite{Li2005}. This comparison highlights the intrinsic advantage of
SrRuO$_3$ in studying current induced magnetization reversal.

In conclusion, we present strong experimental evidence for a non-trivial current contribution to magnetization reversal and we identify
features consistent with magnetic instability induced by spin-wave excitations.

\section{acknowledgments}
We acknowledge useful discussions with A. Kent and A. Mitra. L.K. acknowledges support by the Israel Science Foundation founded by the Israel Academy of Sciences and Humanities.  J.W.R. grew
the samples at Stanford University in the laboratory of M. R. Beasley.

\end{document}